\documentclass[sigconf]{acmart}
\renewcommand\footnotetextcopyrightpermission[1]{} %
\usepackage{balance}

\makeatletter                   
\def\mdseries@tt{m}             
\makeatother                    
\usepackage[draft=true]{minted} 
\usepackage{color}
\usepackage{hyperref}           
\hypersetup{
    colorlinks=true,
    linkcolor=blue,
    filecolor=red,      
    urlcolor=magenta,
    breaklinks=true,            
}
\usepackage{breakurl}           
  
\usepackage{cleveref}
\usepackage{url}

\usepackage{bm}
\usepackage{dsfont}
\DeclareMathOperator*{\argmax}{argmax}
\usepackage{multicol}
\usepackage{graphicx}
\usepackage{color}
\usepackage{subcaption}
\graphicspath{{./figures/}}
\usepackage{booktabs,siunitx}
\newcommand{\ra}[1]{\renewcommand{\arraystretch}{#1}}
\usepackage{tabularx}
\usepackage{etoolbox}
\usepackage{caption} 
\usepackage{wrapfig,lipsum}
\usepackage{placeins}
\usepackage{float}

\usepackage{xcolor,listings}
\usepackage{textcomp}
\usepackage{minted}
\newmintinline[inlinesql]{sql}{}

\usepackage{xspace}
\newcommand*{\eg}{e.g.,\@\xspace}
\newcommand*{\ie}{i.e.,\@\xspace}
\newcommand*{\cf}{cf.\@\xspace}

\DeclareRobustCommand{\parhead}[1]{\textbf{#1}~}

\AtBeginDocument{%
  \providecommand\BibTeX{{%
    \normalfont B\kern-0.5em{\scshape i\kern-0.25em b}\kern-0.8em\TeX}}}

\begin{document}
\sloppy

\title{DeepSPACE: Approximate Geospatial Query Processing \\with Deep Learning}

\renewcommand{\shorttitle}{DeepSPACE}

\author{Dimitri Vorona}
\affiliation{\institution{Technical University of Munich}}
\email{vorona@in.tum.de}

\author{Andreas Kipf}
\affiliation{\institution{Technical University of Munich}}
\email{kipf@in.tum.de}

\author{Thomas Neumann}
\affiliation{\institution{Technical University of Munich}}
\email{neumann@in.tum.de}

\author{Alfons Kemper}
\affiliation{\institution{Technical University of Munich}}
\email{kemper@in.tum.de}

\renewcommand{\shortauthors}{Vorona et al.}

\begin{abstract}
The amount of the available geospatial data grows at an ever faster pace.
This leads to the constantly increasing demand for processing power and storage in order to provide data analysis in a timely manner.
At the same time, a lot of geospatial processing is visual and exploratory in nature, thus having bounded precision requirements.
We present DeepSPACE, a deep learning-based approximate geospatial query processing engine which combines modest hardware requirements with the ability to answer flexible aggregation queries while keeping the required state to a few hundred KiBs.
\end{abstract}


\maketitle

\section{Introduction}
With the growing amount of the available geospatial data new requirements for the query processing emerge~\cite{DBLP:journals/pvldb/EldawyM17}. The highly variable nature of the information presented in the data requires human supervision to discern interesting patterns. This leads to a processing pattern characterized by a high number of ad-hoc queries which ideally should be answered quickly and cheaply, enabling an interactive mode of exploration.

There are multiple ways to meet this challenge. Specialized geospatial indexes enable throughput of millions of data points per second and scale well with the available parallel CPU cores \cite{DBLP:conf/icde/KipfLPPB0K18}. They still require high-end hardware to store the complete dataset and must provide high availability for the user.

Other systems recognize the trade-off between the available hardware and required precision and provide approximate query results based on (online or offline) samples. Such approaches often deliver poor results under very selective queries since the sample size needs to grow inversely with the selectivity to provide the same precision.

In this work we present an alternative approach: \textbf{DeepSPACE} (\textbf{Deep} Geo-\textbf{Sp}atial \textbf{A}utoregressive \textbf{C}onditional \textbf{E}stimator), a compact model which captures the distribution of the data and allows the user to discern interesting patterns while keeping the computation requirements to a level provided even by leanest of today's client devices like smartphones.

We present an unsupervised versatile training regime which works on any kind of geospatial data enriched with additional information about the data points. Using the example of New York City taxi data~\cite{nycopen:online} we show that our model can answer typical exploratory queries with reasonable precision, enabling visual data exploration without the need for fast hardware or connectivity to a backend server.

We base our model on the recent advances in neural distribution estimation~\cite{DBLP:conf/icml/GermainGML15}. Using this type of neural networks opens possibilities to calculate arbitrary conditional distributions effectively providing a query interface for the underlying data.

We use space-filling curves to discretize the geographic locations to facilitate the learning and the querying of the data distributions. In effect, we partition the space using a quad-tree which allows the model to capture the spatial locality relationship of the conditional distributions.

Our model is flexible in regards to input data types and distributions. The modular system allows easy adaptation to new datasets. The user can incorporate their domain knowledge by specifying custom distribution families for specific attributes, \eg a Gaussian mixture for a continuous attribute which is known to be multimodal.

This paper makes the following contributions: (1) we introduce a deep-learning-based approach to approximately answer geospatial aggregation queries which enables interactive data exploration, (2) we show how our architecture can handle heterogeneous queries and data and allow user to extract rich information about the underlying data distributions, and (3) we compare our approach to sample-based baselines and show that we are competitive in the precision of query results even at high sample sizes.

The rest of the paper is structured as follows: Section~\ref{sec:relatedwork} gives an overview over the related work in the areas of geospatial and approximate query processing. In Section~\ref{sec:background} we provide the necessary background to understand the DeepSPACE architecture, which is then described in detail in Section~\ref{sec:deepspace}. We evaluate our system's performance in Section~\ref{sec:evaluation}. We conclude with a summary and overview of future research topics in Section~\ref{sec:summary}.

\section{Related Work}
\label{sec:relatedwork}

Use of approximate query processing in domains that have relaxed requirements to the precision of the results (\eg visualization) is a popular topic in database research. The aim is to provide the answer for arbitrary queries in lower (ideally bounded) time. A recent overview of the modern development in this area is presented in~\cite{DBLP:journals/dase/LiL18}.

Two general approaches to approximate query processing (AQP) are (1) online sampling and (2) offline synopses generation.

Since geospatial queries tend to be very selective, the unbiased approach (\ie based on uniform sampling from the total data set) has to combat the 0-tuple problem (\ie no sample tuples remain after the selection has been applied).

On the other hand, some systems opt for pre-computing a succinct synopsis of the data which can be used to efficiently approximate query results. Wang et al.~\cite{Wang:2015:SOS:2850583.2850584} present an index structure which is enriched with a random sample of points that qualify the filter conditions of the previous queries. Such index can subsequently be used to answer approximate queries efficiently.
In contrast, our approach does not require access to previous queries and provides accurate results early on.

\paragraph{Model-Based Processing.}
There has been some early work on model-based data processing using classical statistical models.
Deshpande et al. suggest to support model-based views (\eg for regression) in a database system~\cite{DBLP:conf/sigmod/DeshpandeM06}.
In contrast to our work, their focus is on the smoothing our the irregularities on the incoming data using simple, easily computable models.

With the mainstream availability of deep neural networks, there has been a recent surge in the number of model-based methods for approximate query processing.
Kulessa et al.~\cite{DBLP:journals/corr/abs-1811-06224} use a sum-product-network-based model to estimate aggregate query results. The authors explore both a direct conditional evaluation, similar to our approach, as well as generation of stratified samples. The extension of their model to geospatial data and comparison to our approach might be an interesting future research topic.

Similarly, Thirumuruganathan et al.~\cite{DBLP:journals/corr/abs-1903-10000} use variational autoencoders to generate samples from the learned joint data distribution. In our context, this approach would suffer from the low selectivity of the typical geospatial queries, since it does not support generation of stratified results.

Yang et al.~\cite{yang2019selectivity} propose the use of an autoregressive model to generate stratified samples and also supports range predicates but does not address geospatial predicates.

\paragraph{Spatial Processing with Bounded Precision.}
Another direction is to trade precision for evaluation performance while using the total dataset.
Kipf et al.~\cite{DBLP:conf/icde/KipfLPPB0K18} use hierarchical grid approximations of polygons to achieve a throughput of tens of millions data points on a single core.
While this approach can guarantee a user-defined precision, it uses a main-memory resident index structure which grows in size with the amount of polygonal data and thus is not directly comparable to this work.

Zacharatou et al.~\cite{DBLP:journals/pvldb/ZacharatouDASF17} propose to leverage the rendering pipeline of the GPU to provide interactive response times over large datasets, optionally with approximate but bounded precision to achieve response times of around one second for almost one billion data points.
In contrast to this approach, ours does not strictly require expensive graphics hardware and only has a modest memory footprint.

\section{Background}
\label{sec:background}

\begin{figure*}[h]
\begin{subfigure}{0.49\textwidth}
\begin{center}
\includegraphics[width=0.8\linewidth]{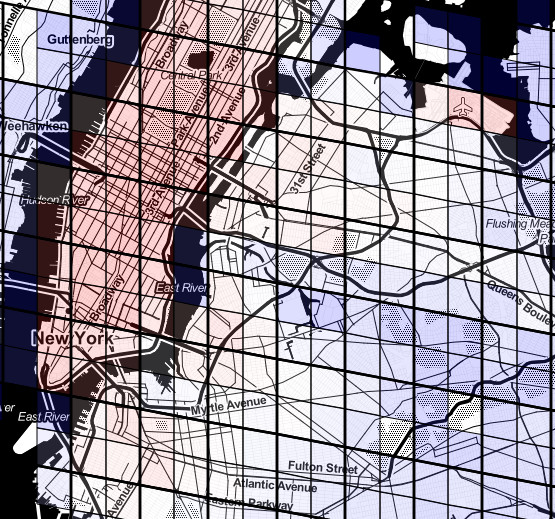}
\end{center}
\caption{Cell grid of levels 12 (thick lines) and 13 (thin lines) superimposed on Manhattan. The number of the taxi rides originating from the specified cell is color-encoded.}
\end{subfigure}
\begin{subfigure}{0.49\textwidth}
\begin{center}
\includegraphics[width=0.4\linewidth]{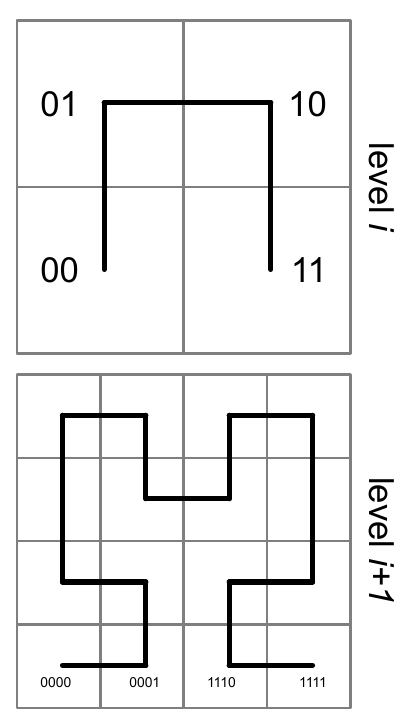}
\end{center}
\caption{Multi-level discretization of a cell using a Hilbert curve.}
\end{subfigure}
\caption{Encoding of geospatial data using space-filling curves.}
\label{fig:geo-background}
\end{figure*}
We base our approach on a number of advances in the areas of geospatial location encoding and distribution modeling. In the following, we provide the necessary background to understand the DeepSPACE architecture.

\subsection{Location Discretization}
\label{sub:loc_discr}

Space-filling curves are a popular approach to location discretization. The idea is to project the geographical surface on a 2d Euclidean surface, split it into nested cells and enumerate the cells according to their position on the space-filling curve. The choice of a Hilbert curve yields a nested cell structure similar to a quad-tree which achieves the best locality preservation. Other options (\eg Z-curve) trade locality for computational efficiency during the computation of the projection.

Figure \ref{fig:geo-background} (left) shows how the discrete cells can be used to visualize geospatial data. The cells at level 14 (which corresponds to approximately $0.32km^{2}$) are outlined in red. Each cell of level 14 consists of four sub cells of level 15. The distribution of the data for each cell (in this case number of rides originating from the area) can be color-encoded on the map.

The hierarchical encoding of the cells is presented in detail in Figure~\ref{fig:geo-background} (right). Each cell consists of up to 30 levels, each level encoded by a number from $0$ to $3$ denoting the cell's position along the Hilbert curve. Each subsequent level is completely covered by the previous level, so the \textit{contains} relation can be checked by simply comparing the prefixes of the cells.

The hierarchical structure allows a natural interpretation as a conditional distribution: each subsequent level can be seen as a categorical (Multinoulli) distribution conditioned on the joint distribution of the preceding cells. This allows the model to exploit the locality properties of the data and improves training.

In this work, we choose the Google S2 library~\cite{S2-lib} as an implementation of the location discretization approach. Besides providing industry-strength stability, S2 is also used by a number of popular database systems (\eg MemSQL~\cite{memsql} and MongoDB~\cite{mongodb}) which allows easy interoperability between these systems and DeepSPACE.

\subsection{Autoregressive Autoencoders}
\label{sub:autoencoders}

Autoencoders~\cite{Goodfellow-et-al-2016} are a type of unsupervised deep learning architecture which aims to learn an efficient encoding for the given data. Autoencoders have many uses, one particular is neural density estimation.

The output of a standard autoencoder does not represent a valid probability distribution. This can be easily seen by considering a ``perfect'' autoencoder: if we interpret its loss as log-likelihood, we get an implied probability density of 1 at every point, which does not integrate to 1, as required for a valid probability distribution. This can be alleviated by enforcing the \textit{autoregressive property}~\cite{DBLP:conf/icml/GermainGML15}. The autoregressive property holds if each output $\hat{x}_i$ only depends on the inputs $\boldsymbol{x}_{0:i-1}$. By enforcing this property we get a valid probability distribution over the input $\boldsymbol{x}$:

$$ p(\boldsymbol{x}) = \prod_{i}p(\hat{x}_i|\boldsymbol{x}_{0:i-1}) $$

The model parameters then can be trained using maximum likelihood estimation.

Of particular interest is the efficient implementation of the autoregressive architecture using masks~\cite{DBLP:conf/icml/GermainGML15}. The idea is to assign each neuron in the layers an index $0..n$ with $n$ being the number of inputs. Then a neuron is connected to a neuron in the previous layer if its index is greater-equals (or strictly greater for the output layer) than the index of the neuron in the previous layer (see also Figure~\ref{fig:blocked_made}). By enforcing the ``strictly greater'' relationship we make sure that inputs of an attributes $0..i$ are not used for the prediction of the output $i$, which corresponds exactly to the \textit{autoregressive property} as defined earlier.

Such connectivity restrictions can be efficiently encoded using a binary mask. The output of a layer thus becomes

$$\phi((W \odot A)x + b) $$

with $\phi$, $W$, $A$, and $b$ being activation function, layer weights, mask, and biases, correspondingly and $\odot$ being Hadamard (elementwise) product of the matrices.

The indexes of the input and output layers arise from the input ordering. The indexes of the inner layers can be assigned arbitrarily from the closed interval $0..n-1$ (since the n-th \ie last input is not used for the output prediction to maintain the \textit{autoregressive property}). This allows to adjust the width and the depth of the model according to the complexity of the data distribution.

\section{DeepSPACE}
\label{sec:deepspace}

In the following, we introduce the DeepSPACE architecture which is able to utilize succinct pre-computed synopses to approximately answer a variety of geospatial queries. DeepSPACE supports a various data types, and can answer common aggregation queries with selection predicates typical for the geospatial domain.

\subsection{Model Architecture}

DeepSPACE utilizes the autoregressive architecture to model a series of conditional distributions which describe the data. In particular, we build on the ideas of masked autoencoders~\cite{DBLP:conf/icml/GermainGML15} to provide an efficient implementation for training and inference.

We extend the autoregressive autoencoder with the notion of \textit{sub-net modules}. A \textit{sub-net module} $\phi_i$ can be considered a function which takes a subset of \textit{input attributes} $x_{0:i-1}$ and produces the parameters of the \textit{target output distribution} $\hat{x}_i$. The input attribute $x_{i}$ is a contiguous set of neurons which describe the input of a data type. 

As an example, let us consider a sub-net $\phi_2$ with a discrete input attribute $x_0$, and a continuous one $x_1$. $x_0$ which consists of $K$ categories would typically be described by a set of $K$ neurons which one-hot encode its value. $x_1$ would have a single input which is simply its value. If the target output distribution of $\phi_2$ is Gaussian, which is common for continuous attributes, then the output parameters consist of two neurons: $\mu$ and $\sigma$, which are the mean and the standard deviation of the Gaussian, respectively.

A sub-net module $\phi_i$ is fit using maximum likelihood estimation on the input attribute $x_i$. Continuing the example above, we search output parameters $\mu$ and $\sigma$ such that

$$ \argmax_{\mu,\sigma}\mathcal{L}_{\text{Gaussian}}(\mu, \sigma | x_{i}) $$

The changes are then applied to the hidden units of the subnets using back-propagation. 

Alternatively, a single sub-net module can be seen as an independent neural network or any other function approximator. In other contexts, this view on autoregressive models is taken by other authors (\cf \cite{DBLP:conf/nips/PapamakariosMP17,blockneural_decao,yang2019selectivity}).

\begin{figure*}[h]
\centering
\def\svgwidth{0.65\textwidth}
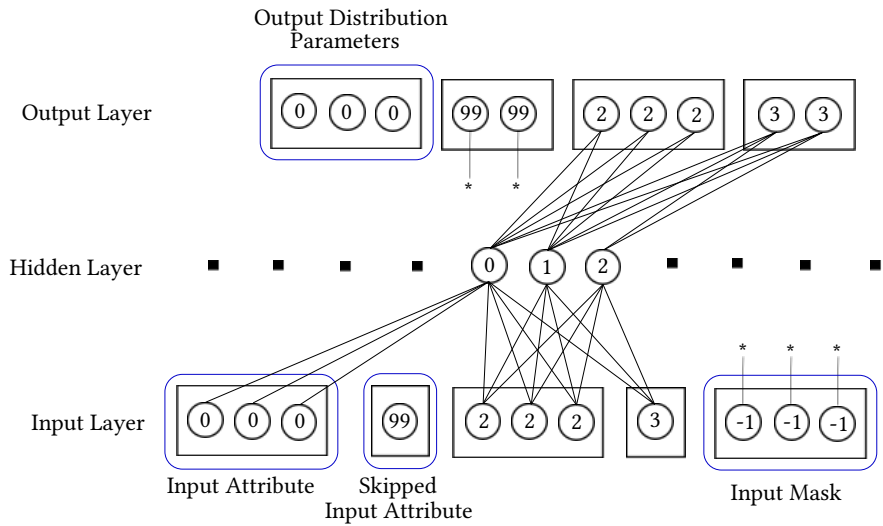
\caption{DeepSPACE's autoregressive architecture. The hidden layer neurons are given order numbers which are used to decide which inputs and outputs are connected to them. The lines denote connections to individual neurons, * represents a connection to every neuron in the neighboring layer.}
\label{fig:blocked_made}
\end{figure*}

DeepSPACE embeds the sub-net modules in a single neural network as pictured in Figure~\ref{fig:blocked_made}. Input and output neurons of a subnet are labeled with the subnet's index. Special inputs and outputs are given placeholder values which are smaller or larger than any neuron index which forces a full or prevents any connection to the neighboring layer, respectively. One use of it is to encode an input mask which describes which input attributes are used for predictions. The attributes not used as input are not connected to any neurons, the corresponding output distributions, on the other hand, are connected to every output except itself.

The neurons in the hidden layer can be assigned indexes in an arbitrary fashion, though this assignment influences the participation of a single neuron in the sub-net modules. The original MADE authors suggest using uniform random assignment~\cite{DBLP:conf/icml/GermainGML15}. We, on the other hand, assign the indexes proportionally to the number of the sub-nets a neurons participates in, \ie the number of neurons with an index $i$ is proportional to $i$.

We then enforce the autoregressive property as described in Section~\ref{sub:autoencoders}. The architecture can be trivially extended to multiple hidden layers by inserting additional hidden layers and assigning indexes to the neurons as described above.

\subsection{Order-Agnostic Attribute Ordering}
\label{sub:orderagnostic}

The presented architecture imposes an ordering on the input attributes. This means that $\phi_i$ cannot be evaluated independently of the input attributes $j$ for $j < i$. While, due to the product rule, any ordering of the input attributes should return the same result, we must be able to ``skip'' certain attributes if we are not interested in an output distribution conditional on them. One approach is to marginalize the attribute by either evaluating every possible value and summing the results or by applying a Monte Carlo approach if the number of distinct values is intractable~\cite{yang2019selectivity}. The computation requirements grow linearly with the number of distinct categories or the size of the Monte Carlo sample, respectively.

DeepSPACE instead utilizes the training techniques from~\cite{DBLP:conf/icml/UriaML14} to 
create an \textit{order agnostic} representation of data. We organizes the connections from the input layer and to the output layer depending on the which attributes are used as conditional inputs and which are used as outputs. Concretely, we set the indexes of the ``skipped'' input neurons and the output-only neurons to a placeholder value higher than any other neuron index, thus enforcing the desired connectivity pattern (\cf Figure~\ref{fig:blocked_made}). An additional advantage of this approach is that we are able to get an estimation for multiple output distributions in a single pass.

To facilitate model training, we provide an additional input mask which denotes the attributes on which the filters are defined (\cf~\cite{DBLP:conf/icml/UriaML14}). This mask is connected to each layer (\cf Figure~\ref{fig:blocked_made}) and thus provides every neuron with the information on conditional input and output attributes. With this architecture, we are able to answer conjunctive equality predicates using a single forward-pass on our model.

By combining different orderings of sub-net modules into a single neural network, we not only improve evaluation performance, but also allow information sharing across the sub-nets, since the individual sub-nets partially share neurons.

\subsection{Handling Heterogeneous Data}
\label{sub:handlingdata}

DeepSPACE can handle arbitrary data distributions by encoding their parameters in the corresponding input and output blocks. In the following, we describe our encoding scheme for \textit{categorical}, \textit{datetime}, \textit{geospatial}, and \textit{continuous} data types.

\parhead{Categorical data} with $K$ distinct categories is encoded using $K$-ary one-hot input vectors. The output parameters are logits (log odds) of the possible classes which can be converted to probabilities by applying the \textit{softmax} function $\sigma(x)_i = \frac{exp(x_i)}{\sum_j{exp(x_j)}} $. Conversely, we use the categorical cross-entropy loss function to maximize the log-likelihood of the sub-net module:

\begin{equation*}
\begin{split}
\text{loss} (\phi_{\text{cat},i}(x_{0:i-1})) & = \text{cross entropy}(x_i, \phi_{\text{cat,i}}(x_{0:i-1})) 
\\ & = -\log\left(\frac{\exp(\phi_{\text{cat},i}(x_{0:i-1})[x_i])}{\sum_j \exp(\phi_{\text{cat},i}(x_{0:i-1})[j])}\right)
\\ & = - \phi(x_{0:i-1})[x_i]) + \sum_j{\phi_{\text{cat},i}(x_{0:i-1})[j]}
\end{split}
\end{equation*}
where $v[i]$ denotes the $i$th output of the one hot-encoded vector $v$.

A binary encoding of the data (\cf~\cite{DBLP:journals/corr/abs-1903-09999}) would yield a more compact model, but lead to less easily interpretable output and require a more complex model. One reason for this is that the binary encoding implies relationships between categories which do not have an equivalent in the data, \eg the category $00$ is equally related to $11$ as to $01$, but the input format suggests otherwise.

\parhead{Datetime} can be seen as a continuous attribute \textit{time since epoch} or a discrete attribute encoding a value extracted from the continuous timestamp, \eg \textit{month}, \textit{day of the week}, or \textit{hour}. Since a typical selection or group-by query targets datetime data types on higher granularity and discrete encoding facilitates the training of the model, we choose the discrete encoding in DeepSPACE. However, sub-net modules can model arbitrary distributions, so a continuous encoding for the datetime data type can be added without any architectural changes.

\parhead{Geospatial data} presents a special case of categorical data. As described in Section~\ref{sub:loc_discr}, the geospatial area is split using a quad-tree-like schema into ever finer segments. We consider each split at the level $L$ to be a categorically distributed variable $G_L$ with the number of categories $K = 4$ (\ie $G_L \sim Cat(4)$). The splits on the next level can then be modeled as random variable conditionally dependent on the previous levels $G_{L+1} \sim Cat(4) \times G_{L}$. This fits neatly into our autoregressive architecture. The loss function for each layer is the categorical cross entropy used for categorical data.

Discrete encoding of the geographical input data provides a number of distinct advantages over continuous latitude and longitude inputs. First, it provides a strong inductive bias for the model by only allowing the distributions of the higher levels to affect the distributions of the lower levels. Second, it splits the complex multi-modal joint distribution into an easy-to-model set of categorical distributions. Third, handling categorical distributions is more numerically stable than using IEEE 754 float encoded geographic coordinates.

The nested discretization schema captures the local relationships of the data and allows efficient querying on regions of diverse sizes. Alternative encoding schemes like binary or Z-curve-based encoding, while feasible, are at a disadvantage in one of these areas.

\parhead{Continuous data} is encoded using a diagonal mixture of $N$ Gaussian distributions, or for less complex data a single Gaussian, which corresponds to $N=1$. The input is a single real value, the output is a vector of triples $(a_i, \mu_i, \sigma_i)$ of mixture factor, mean, and standard deviation, correspondingly. In practice, we output $log(\sigma)$ to ensure a positive $\sigma$ value. Since we assume diagonal (\ie mutually uncorrelated) Gaussians, it is sufficient to learn a single standard deviation per mixture component.

It also can be useful to fit the Gaussians on logarithms of the attribute values. This corresponds to learning of a log-normal distribution with the parameters $\mu$ and $\sigma$.

The sub-network $\phi_{\text(cont),i}$ is subsequently fitted using the log-likelihood maximizing loss function: 

\begin{equation*}
\begin{split}
\text{loss}(\phi_{\text{cont},i}(x_{0:i-1}) & = -\sum_{i=0}^{N-1}{log[a_i\mathcal{N}(x|\mu_i,\sigma_i)]} \\ & = -\sum_{i=0}^{N-1}{[log(a_i) + log(\sigma) -\frac{1}{2\sigma^2}(x_i - \mu)^2]}
\end{split}
\end{equation*}

The decision of the number of Gaussians in the mixture is a hyperparameter which has to be chosen by an expert with the domain knowledge based on the complexity of the targeted attribute data. In our experience with the NYC taxi data, a single Gaussian fitted on an attribute or its logarithm is often enough to capture the distribution with enough precision for approximate query processing.

Some attributes, particularly financial data like individual wealth, are better described using the \textit{Pareto distribution} distribution given by $\text(Pareto(x|\alpha,\beta) = \alpha\beta^{\alpha}x^{-\alpha+1}$ for $x\geq\beta, \alpha,\beta>0$ (see~\cite{DBLP:books/lib/Murphy12}, Chapter 2.4.6). By assigning $\beta$ some small positive value known to be less than every $x$, the output parameter of the sub-net modules to $log(\alpha)$ (to ensure its positivity), and slightly adjusting the loss function, we can fit our parameters to this distribution\footnote{In practice one would use some smooth variant of the Pareto distribution, like hybrid Pareto~\cite{hybrid_pareto_carreau}, but we omit its details for brevity.}:

\begin{equation*}
\begin{split}
\text{loss}(\phi_{\text{pareto},i}(x_{0:i-1})) & =  \log(\alpha\frac{\beta^{\alpha}}{{x_i}^{\alpha+1}})  \\ &=  \log(\alpha) + \alpha \log(\beta) - (\alpha+1)\log(x_i).
\end{split}
\end{equation*}

This demonstrates the flexibility of our approach: any distribution can be modularly plugged in into the architecture as a sub-net by defining its inputs, output parameters, and the log-likelihood-based loss function.

\subsection{Query Processing}

Our model can answer a variety of conjunctive queries typical in a geospatial setting. In the following, we describe our approach to processing the queries using our architecture.

A single selection operation of applying a predicate $Pred$ on an attribute $X_i$ can be interpreted as imposing the evidence 
$ P(x | x \in X_i \land Pred(x)) $ 
on the joint prior $ P(\boldsymbol{X}) $,  
\ie $ P_{selected}(\boldsymbol{X}) \sim P(x | x \in X_i \land Pred(x)) \times P(\boldsymbol{X}) $. In other words, $ P(x | x \in X_i \land Pred(x)) $ is the proportion of the tuples which qualify for the predicate $Pred(x)$, $P(x)$ is the data distribution before applying the filter.

Using the product rule, we can recursively generalize this to $N$ selection operations:

$$ P_{selected, N}(\boldsymbol{X}) = \prod_{i=1}^N{P_{selected, i-1}(x | x \in X_i \land Pred(x))}$$

with $ P_{selected,0} = 1 $. The distribution $ P_{selected,i} $ is exactly the conditional distributions learned by the sub-net module $\phi_i$.

To provide an example, we first look into evaluation of an equality predicate on a categorically distributed attribute $j$. To acquire an estimation of the distribution parameters of a target column $i$, we estimate the probability mass $\phi_{j}(x_{j})=\sum_{k}{phi_{j}(k)\llbracket k=j\rrbracket}$ with $x_{j}$ and $\llbracket k=x \rrbracket$ being the filter value and Iverson bracket, respectively. Then take a product with the target density function $\phi_{i}$. The target distribution in this case can be any continuous or discrete distribution, as described in Section~\ref{sub:handlingdata}. The column $j$ is also marked with a $1$ in the input mask to indicate it taking part in the conditional evaluation (\cf Section~\ref{sub:orderagnostic}).

We support filter predicates on any attribute subset, as described in Section~\ref{sub:orderagnostic}. The output distributions, which can be roughly thought of as aggregations on columns in the \lstinline{SELECT} clause of a SQL query, provide basis for the evaluation of the aggregations. For example, a sum can be thought of as the product of the number of the elements in the selected region and the mean of the fitted distribution. More advanced aggregation functions like \lstinline{PERCENTILE} can be supported similarly. Aggregated views like a histogram can also be easily extracted from the distribution, while such queries are not straight-forward in conventional query processing systems.

Since geospatial data is a special case of categorical data, it is handled in a similar fashion. When a query defines a selection of an area described by a specific cell, we apply a selection sequentially on each level up to the targeted one (top down). Unlike in the case of categorical attributes, cell layers cannot be ``skipped'' in a query, though the maximal cell depth can vary. We handle it by extending the input attribute mask by the number of cell levels and setting all values to $1$ up to the position that corresponds to the level of the selection cell.

To handle selections on more complex geospatial shapes (\ie polygons), we approximate the polygons using a set of cells at a specific level. Smaller cells provide a better approximation of a polygon, but generate a higher number of sub-queries to evaluate. The results of sub-queries are then combined depending on the type of the queried aggregate.

For attributes with a limited number of distinct values (up to a few thousand), range predicates can be computed by marginalization. For categorical attributes with higher cardinalities or continuous attributes, we can use the recently proposed \textit{progressive Monte Carlo sampling}~\cite{yang2019selectivity}. This approach generates samples using the conditional information from the previous attributes and thus avoids the necessity of generating an exponential number of samples to achieve a specified precision with a growing number of filtered columns.

\subsection{Aggregates Computation}

DeepSPACE supports the following aggregation functions: \textbf{COUNT}, \textbf{MEAN}, \textbf{STDDEV}, \textbf{PERCENTILE}, \textbf{SUM},  \textbf{MIN}, and \textbf{MAX}. We describe briefly how their computation is implemented on continuous data types modeled by a single Gaussian, and skip the trivial case of categorical data and the more complex case of Gaussian mixtures for brevity:

\begin{itemize}
    \item \textbf{COUNT} is computed by multiplying the estimated proportion of the qualifying tuples with the total number of the entries. 
    \item \textbf{MEAN} and \textbf{STDDEV} for continuous attributes are simply the parameters $\mu$ and $\sigma$ of the fitted Gaussian.%
    \item \textbf{PERCENTILE} is computed using the CDF of the Gaussian: $\frac{1}{2}[1+erf(\frac{x-\mu}{\sigma\sqrt{2}})]$
    \item \textbf{SUM} is the product of count and mean.
    \item \textbf{MIN} and \textbf{MAX} are computed using the following approximation~\cite{gaussian_minmax} 
    $$E[r,n]\approx\mu+\sigma+\Phi^{-1}(\frac{r-\frac{\pi}{8}}{n-\frac{\pi}{4}+1})$$
    
    where $r$ is the desired rank of the value and $n$ number of samples, \ie $n=\text{\textbf{COUNT}}$, $r=1$ for \textbf{MIN} and $r=n$ for \textbf{MAX}.
\end{itemize}

\subsection{Model Training}
\label{sub:model_training}

We train the model using gradient descent on the mini-batched data input with the goal of the minimization of the negative log-likelihood across the sub-nets. To achieve order-agnostic training we have to consider every acceptable input ordering, \ie every subset of the attributes and every cell depth for every mini-batch. Since the number of combinations grows with the factorial, we sample a combination of filter attributes and normalize the loss as described in \cite{DBLP:conf/icml/UriaML14}.

To maintain numerical stability while dealing with low selectivities, we evaluate the conditional probabilities in log space. %

Since we aim to capture the data distribution as precisely as possible, we do not apply any regularization techniques.

\section{Evaluation}
\label{sec:evaluation}

To assess the performance of DeepSPACE, we evaluate it on a number of common tasks and compare with a sampling-based approach.

\subsection{Dataset}
\label{eval:dataset}

We choose the publicly available New York City (NYC) taxi rides dataset~\cite{NYCTaxiCommission2019} for the evaluation. This dataset consists of approximately 350,000 taxi rides per day, containing the information about date, time, and location of the origins and destinations of the taxi rides, as well as additional data like taxi fare, number of passengers, and the tip received by the driver.

We train a DeepSPACE model on the data from January 2016 which consists of around 10 million records. We also trained a separate model on the combined data from January and February to test the ability of the model to exploit regularities in the data.

To demonstrate the versatility of our approach we choose \textit{day of month}, \textit{day of week}, and \textit{hour} as datetime, \textit{pickup location} as geospatial, and \textit{total fare} as continuous attributes. See Section~\ref{sub:handlingdata} for details on encoding of the data types.

\subsection{Model Hyperparameters}

The DeepSPACE model can be trained with arbitrary depth and width of hidden layers. We choose two hidden layers of 386 exponential linear units (ELU)~\cite{DBLP:journals/corr/ClevertUH15}) each. This corresponds to a model size of around 520\,KiB. The model is trained using the Adam optimizer~\cite{DBLP:journals/corr/KingmaB14} with a learning rate of $10^{-4}$ on mini-batches of size 1024. The model is trained for 300 epoch using early stopping, if no improvement was observed for the last 20 epochs as measured on the negative log-likelihood loss on the training data. The actual training time depends on the employed hardware and the dataset size, and it benefits greatly from the acceleration on modern GPUs.

We do not perform model validation since we are not concerned with the generalization capabilities of our model.

\subsection{Geospatial Query Processing}

As the first experiment, we evaluate the precision on the queries of the format 

\begin{minted}{sql}
SELECT COUNT(*) 
    FROM yellow_taxi 
    WHERE ST_CellContains(<cell_id>, pickup_location)
\end{minted}

where \lstinline{ST_CellContains} is defined similar to \lstinline{ST_Contains} to return true when the point given as the second argument is located inside the S2 cell given as the first argument. This kind of queries is natural for the generation of geospatial visualizations as discussed in Section~\ref{sub:loc_discr}. We randomly select a cell at a level between 13 and 16 (corresponding to side lengths of approximately 1,150 to 140 meters~\cite{S2-lib}) from the NYC taxi data area which contains at least a single data point.

\newcommand{\mc}[3]{\multicolumn{#1}{#2}{#3}}

\begin{table*}
\sisetup{detect-weight,mode=text}
\newrobustcmd{\B}{\bfseries}
\ra{1.2}
\centering
\setlength{\tabcolsep}{0pt}
\begin{tabular*}{\textwidth}{
  @{\extracolsep{\fill}}
  p{2cm}
  cccc
  cccc
  cccc
}
\toprule
& \mc{3}{c}{Level $13$ [$1.27\text{ km}^2$]} & \mc{3}{c}{Level $14$ [$0.32\text{ km}^2$]} & \mc{3}{c}{Level $15$ [$0.08\text{ km}^2$]} & \mc{3}{c}{Level $16$ [$0.02\text{ km}^2$]}\\
\cmidrule{2-4} \cmidrule{5-7} \cmidrule{8-10} \cmidrule{11-13}
& {$\varnothing$} & {50th} & {95th} 
& {$\varnothing$} & {50th} & {95th}
& {$\varnothing$} & {50th} & {95th}
& {$\varnothing$} & {50th} & {95th} \\
\midrule
\textbf{DeepSPACE} &   1.20 & 1.08 &  1.87 &   1.16 &  1.09 &    1.55 &   1.16 &   1.09 &  1.52 &   1.18 &   1.10 &  1.57 \\

sample 0.1\% &  153 &  2.28 & 652 & 184 &  1.93 &  914 &  188 &  1.59 &   991 & 183 & 1.54 & 1100 \\

sample 1\% & 9.89 & 1.17 & 3.4 & 11.8 & 1.14 & 4.05 & 3.95 & 1.12 & 2.32 & 2.46 & 1.13 & 1.93 \\

sample 10\% & \B 1.10 & \B 1.04 & \B 1.41 & \B 1.09 & \B 1.04 &\B 1.33 &  \B 1.07 &   \B 1.03 & \B 1.27 & \B 1.06 & \B 1.04 & \B 1.20 \\
\bottomrule
\end{tabular*}
\caption{Q-errors on the query workload for the \textbf{COUNT(*)} aggregate and a predicate on the geospatial location. The results are grouped by the cell level of the filter predicate.}
\label{tab:resultsgeo}
\end{table*}

We compare the model performance with a sampling-based approach with different sample sizes on the q-error metric~\cite{DBLP:journals/pvldb/MoerkotteNS09}. The q-error [$q \geq 1$] is the factor between the estimated and the true cardinality (or vice versa). Alternatively, the q-error can be thought of as the measure of the absolute difference between the estimation and the true result in the log-space, \ie $exp(|log(x_{est}) - log(x_{true})|)$. Since the q-error is undefined for true or estimated cardinalities of zero (latter being a common case in sampling-based estimation)m we disregard queries that return a true result of zero. We further interpret zero-sized results from the sampling estimation as one. For DeepSPACE, no such adjustment is necessary. 

Table~\ref{tab:resultsgeo} shows the results. DeepSPACE provides better performance than the sampling-based approaches up to sample size of 10\%. In absolute numbers, DeepSPACE provides query results within 10\% of the true values in median for all cell levels. While the sample size of 1\% provides an acceptable median precision, the average performance suffers from the minority of cases where the sample only captures a very small number of data entries.

\begin{table}[H]
\ra{1.2}
\centering
\setlength{\tabcolsep}{0pt}
\begin{tabular}{
  @{\extracolsep{\fill}}
  p{7.0cm}
  c
}
\toprule
& State Size \\
\midrule
\textbf{DeepSPACE} & 520\,KiB \\
\textbf{sample 0.1\% ($\sim10,000$ entries)} & 160\,KiB \\
\textbf{sample 1\% ($\sim100,000$ entries)} & 1.6\,MiB \\
\textbf{sample 10\% ($\sim1,000,000$ entries)} & 16\,MiB \\
\bottomrule
\end{tabular}
\caption{State size of the different estimation approaches.}
\label{tab:statesize}
\end{table}

For reference, we provide the state sizes required by each estimation method in Table~\ref{tab:statesize}. DeepSPACE provides vastly better estimations than the 1\% sample while consuming only one third of the space. Note that we do not employ any compression techniques for the samples.

\subsection{Additional Predicates}
\label{sub:addpred}

Next, we add filter predicates on the datetime columns, \ie limiting the queried information to a certain date, date of the week, or hour.

For the additional predicates, we randomly choose between 1 or 2 columns, taking care not to include \textit{day of month} and \textit{day of week} at the same time, since the latter is determined by the former in a single month. An example SQL query would be

\begin{minted}{sql}
SELECT COUNT(*) 
    FROM yellow_taxi 
    WHERE ST_CellContains(<cell_id>, pickup_location)
          AND EXTRACT(DAYOFWEEK FROM 
                pickup_time) = <day_of_week>
\end{minted}

We expect this to model the natural exploratory query workload where the user is interested in a subset of information limited by the time and area. The resulting selectivities thus can be considered typical for the geospatial domain.

\begin{table*}[h]
\sisetup{detect-weight,mode=text}
\newrobustcmd{\B}{\bfseries}
\ra{1.2}
\centering
\setlength{\tabcolsep}{0pt}
\begin{tabular*}{\textwidth}{
  @{\extracolsep{\fill}}
  p{2cm}
  ccc
  ccc
  ccc
}
\toprule
& \mc{3}{c}{$N < 100$} & \mc{3}{c}{$100 \leq N < 1000$} & \mc{3}{c}{$N \geq 1000$}\\
\cmidrule{2-4} \cmidrule{5-7} \cmidrule{8-10}
& {$\varnothing$} & {50th} & {95th}
& {$\varnothing$} & {50th} & {95th}
& {$\varnothing$} & {50th} & {95th} \\
\midrule
\textbf{DeepSPACE} &   \B 1.83 & \B 1.42 &  \B 3.95 &   1.21 &    1.14 &    1.61 &   1.10 &   1.07 &  1.28 \\
sample 0.1\% &   23.8 &   13.0 &   81.0 &   228 &   171 &    718 &   256 &   1.53 &   1600 \\
sample 1\% & 16.3 & 7.00 &   66.0 & 22.3 &   1.51 &   158 &  1.22 &   1.14 &  2.25\\
sample 10\% & 3.18 & 1.69 & 11.0 & \B 1.19 & \B 1.13 &\B 1.54 &  \B 1.06 &   \B 1.04 & \B 1.16 \\
\bottomrule
\end{tabular*}
\caption{Q-errors on the query workload for the \textbf{COUNT(*)} aggregate, predicates on geospatial location and the datetime columns. The result are grouped by the number of the entries $N$ which qualify the predicates.}
\label{tab:resultscount}
\end{table*}

Table~\ref{tab:resultscount} presents the results. The results are grouped based on query selectivity. The most selective group ($N<100$) represented 64\% of queries and had median selectivity of $1.19*10^{-6}$. At this selectivity our model provides better performance than sampling 10\% of the data or less. The queries with results between 100 and 1000 constituted 26\% of the workload (median selectivity $2.50*10^{-5}$. At this group our model loses slightly to sampling 10\%. A similar picture presents itself in the least selective group ($N\geq1000$, median selectivity $1.79*10^{-4}$) in which 10\% of the queries fell. Over the whole query workload, DeepSPACE offered the best performance with a median q-error of 1.25 compared to 1.28 of the sampling of 10\% of the datasets (1.60 versus 2.46 in mean).

\subsection{Effect of Dataset Size}

To investigate the effect of the increased dataset size, we the train our model on the data from January and February, effectively doubling it. We keep the absolute sizes of the samples constant which means that the proportion of the sampled entries relative to the dataset size gets halved. To maintain the absolute selectivity of the queries, we add a random equality predicate on the month column to each query and keep the rest of the workload the same.

\begin{table*}
\sisetup{detect-weight,mode=text}
\newrobustcmd{\B}{\bfseries}
\ra{1.2}
\centering
\setlength{\tabcolsep}{0pt}
\begin{tabular*}{\textwidth}{
  @{\extracolsep{\fill}}
  p{2cm}
  ccc
  ccc
  ccc
}
\toprule
& \mc{3}{c}{$N < 100$} & \mc{3}{c}{$100 \leq N < 1000$} & \mc{3}{c}{$N \geq 1000$}\\
\cmidrule{2-4} \cmidrule{5-7} \cmidrule{8-10}
& {$\varnothing$} & {50th} & {95th}
& {$\varnothing$} & {50th} & {95th}
& {$\varnothing$} & {50th} & {95th} \\
\midrule
\textbf{DeepSPACE} &   \B 1.67 & \B 1.37 &  \B 3.21 &  \B  1.26 &  \B   1.16 &  \B   1.76 &  1.11 &   1.08 &  1.32 \\
sample 0.05\% &   28.6 &   19.0 &   85.0 &   332 &   278 &    846 &   445 &   1.57 &  2507 \\
sample 0.5\% & 23.3 & 13.0 &  78.0 & 60.3 &  1.78 &  336 & 2.30 &   1.14 &  1.83 \\
sample 5\% & 6.08 & 2.06 & 25.0 & 1.33 & 1.18 & 1.81 &  \B 1.06 &   \B 1.04 & \B 1.20 \\
\bottomrule
\end{tabular*}
\caption{Q-errors on the query workload for the \textbf{COUNT(*)} aggregate on the doubled amount of data.}
\label{tab:resultstwomonth}
\end{table*}

The results are presented in Table~\ref{tab:resultstwomonth}. Our model performance remains practically unaffected by the increase in the total amount of data. This confirms our conjecture that our model is able to exploit the similarities in the subsequent periods to succinctly summarize the data using constant state size. The performance of the sampling-based approaches on the other hand suffers for the more selective queries due to the fact that a smaller number of qualifying entries get into each sample.

\subsection{Querying Continuous Attributes}

Finally, we evaluate the performance on the continuous attributes using the aggregation functions \textbf{SUM} and \textbf{MEAN} with the same types of predicates as in Section~\ref{sub:addpred}. We use the symmetric mean absolute percentage error (\textit{sMAPE}) to compare the estimation precision. \textit{sMAPE} is defined as 

$$ \sum_{i=0}^{Q}{\frac{\mid S_{i,true} - S_{i,est}\mid}{(S_{i,true} + S_{i,est}) / 2}} $$

where $Q$ is the number of queries, $S_{i,true}$ and $S_{i,est}$ are true and estimated results of $i$-th query, respectively. Unlike the relative error, another popular approximation error metric, \textit{sMAPE} is bounded ($[0, 2]$) and less asymmetric in respect to under- and over-estimations. 

\begin{table*}
\sisetup{detect-weight,mode=text}
\newrobustcmd{\B}{\bfseries}
\ra{1.2}
\centering
\setlength{\tabcolsep}{0pt}
\begin{tabular*}{\textwidth}{
  @{\extracolsep{\fill}}
  p{2.1cm}
  ccc
  ccc
  ccc
}
\toprule
& \mc{3}{c}{$N < 100$} & \mc{3}{c}{$100 \leq N < 1000$} & \mc{3}{c}{$N \geq 1000$}\\
\cmidrule{2-4} \cmidrule{5-7} \cmidrule{8-10}
& {$\varnothing$} & {50th} & {95th}
& {$\varnothing$} & {50th} & {95th}
& {$\varnothing$} & {50th} & {95th} \\
\midrule
\textbf{DeepSPACE} &  \\
\textit{MEAN} & \B 0.23 & \B 0.15 &  \B 0.73 &  \B  0.06 &  \B  0.06 &  \B 0.16 &   0.05 &   0.04 &  0.11 \\
\textit{SUM} & \B 0.46 & \B 0.35 &  \B 1.30 &  \B  0.17 &   \B  0.13 &  \B   0.46 &   0.10 &   0.07 &  0.26 \\
\textbf{sample 1\%} &  \\
\textit{MEAN} & 1.72 & 2.00 &  2.00 &   0.51 &   0.29 &    2.00 &   0.12 &   0.16 &  0.33 \\
\textit{SUM} & 1.81 & 2.00 &  2.00 &   0.69 &    0.50 &    2.00 &   0.21 &   0.16 &  0.57 \\
\textbf{sample 10\%} &  \\
\textit{MEAN} & 0.94 & 0.92 &  2.00 &  0.11 &   0.08 &  0.52 & \B  0.04 & \B  0.00 & \B 0.09 \\
\textit{SUM} & 1.10 & 0.91 &  2.0 &  0.19 & 0.15 &  0.52 & \B 0.06 & \B 0.05 & \B 0.18 \\
\bottomrule
\end{tabular*}
\caption{sMAPE metric for the estimation of the \textbf{MEAN} and \textbf{SUM} aggregates.}
\label{tab:resultssummean}
\end{table*}

For this experiment, we skip evaluation on the sample size of 10 thousand entries since the results were vastly inferior to other approaches.

The results are presented in Table~\ref{tab:resultssummean}. DeepSPACE provides overall best precision for the queries with up to 1000 qualifying entries. Only in the group with more than 1000 entries, the sampling of 10\% of the data provides slightly better estimates.
\section{Summary and Future Work}
\label{sec:summary}

The constantly increasing size of the available geospatial data presents new challenges to query processing systems. We have presented DeepSPACE, a deep learning-based approximate query processing engine which provides the user with quick and precise estimates while requiring only a small state size of few hundred KiBs.

We have introduced a novel modular model architecture which is able to support the most common data types and can be easily extended to support arbitrary data distributions. Our evaluation on the NYC taxi dataset showed that DeepSPACE provides better precision on the query workloads typical for the geospatial domain than our sampling baseline.

In the future, we plan to extend DeepSPACE to support a wider range of datatypes and complex data distributions. A more thorough evaluation on a wider range of datasets could also offer valuable insights into the strengths and challenges of our approach. Finally, we would like to compare our model against more sophisticated competition like sum-product based networks~\cite{DBLP:journals/corr/abs-1811-06224} extended with the support for geospatial data types.

\bibliographystyle{abbrv}
\bibliography{main}

\end{document}